# A counter-argument to "Hidden variable models for quantum theory cannot have any local part"

*Sofia Wechsler* [1)]

**Abstract**
Colbeck and Renner analyzed a class of combined models for entanglements in which local and non-local hidden variables cooperate for producing the measurement results. They came to the conclusion that the measurement results are fully independent of the local components of the hidden variables. Their conclusion is based mainly on an assumption on the local hidden variables, assumption similar to the non-signaling property of probabilities of observables' values.
In the present text it is proved that hidden variables are not observables, so their distributions of probabilities do not necessarily possess the non-signaling property. Also, a counter-example is brought to the Colbeck and Renner assumption, showing that their type of models and conclusion are not general. The question whether hidden variables, local or non-local, exist or not, remains open.

**Abbreviations**
HV = hidden variable
L.H.S. = left hand side
R.H.S. = right hand side

## 1. Introduction and a lemma about hidden variables

Colbeck and Renner [1] analyzed a class of combined models for entanglements in which local and non-local hidden variables cooperate for producing the measurement results. They came to the conclusion that the measurement results are fully independent of the local components of the hidden variables.
Their proof is based mainly on a special assumption that looks like the non-signaling property. The non-signaling property says that given two entangled particles, the probabilities of the values of one particle's *observables*, are independent of the settings at the station where the other particle is tested. However, in [1] this property is extended to hidden variables. It is shown below that a hidden variable is not an observable, so it is not justified to apply the non-signaling property to it.

**Lemma.** *Hidden variables that account for values of non-commuting observables, are not observables.*

Proof: consider an arbitrary state of a single particle moving in the direction *z*, and let *S* be the hidden variable responsible for the position *Z* and for the linear momentum $\Pi_Z$, i.e. $Z = c(S)$, $\Pi_Z = d(S)$. If *S* is an observable then a hermitian operator $\hat{S}$ is associated with it. Let's calculate the commutator $\hat{Z}\hat{\Pi}_Z - \hat{\Pi}_Z\hat{Z} = c(\hat{S})\,d(\hat{S}) - d(\hat{S})\,c(\hat{S})$. A hermitian operator commutes with itself and any two functions of it also commute. Therefore $c(\hat{S})\,d(\hat{S}) - d(\hat{S})\,c(\hat{S}) = 0$. But that is impossible because $\hat{Z}$ and $\hat{\Pi}_Z$ don't commute. Therefore $\hat{S}$ doesn't commute with itself, so it is not hermitian and *S* is not an observable.

The significance of this lemma is that the hidden variables are not physical properties that due to technical limitations of our time escape to our observation. The impossibility to observe the hidden variables is not a technical limitation is it a *no go* condition of the nature.

---

[1)] Computers Engineering Center, Nahariya, P.O.B. 2004, 22265, Israel

## 2. The Colbeck and Renner class of models for entanglements

Consider two photons, *1* and *2*, prepared in the polarization-singlet state

(1) $|\psi\rangle = 2^{-1/2}(|d\rangle_1|d\rangle_2 + |d\perp\rangle_1|d\perp\rangle_2)$,

where $d\perp$ is a direction perpendicular on the direction $d$. The photon *1* flies to the station of the experimenter Alice, and the photon *2* flies to the station of the experimenter Bob. Alice chooses some direction $a$ to test the polarization of the photon *1*, and Bob chooses some direction $b$ to test the polarization of the photon *2*. The direction $a$ makes an angle $\theta$ with the direction $d$ in the plane $d$-$d\perp$, and the direction $b$ makes an angle $\phi$ with $d$ in the same plane. The transformations undergone by the two photons are

(2) $|d\rangle_1 = \cos\theta\,|a\rangle + \sin\theta\,|a\perp\rangle$, $\quad |d\perp\rangle_1 = -\sin\theta\,|a\rangle + \cos\theta\,|a\perp\rangle$, $\quad$ (a)

$|d\rangle_2 = \cos\phi\,|b\rangle + \sin\phi\,|b\perp\rangle$, $\quad |d\perp\rangle_2 = -\sin\phi\,|b\rangle + \cos\phi\,|b\perp\rangle$. $\quad$ (b)

Introducing (2) in (1)

(3) $|\psi\rangle = 2^{-1/2}\{\cos(\phi - \theta)(|a\rangle|b\rangle + |a\perp\rangle|b\perp\rangle) + \sin(\phi - \theta)(|a\rangle|b\perp\rangle - |a\perp\rangle|b\rangle)\}$.

The measurement results are binary and we will denote by $X = +1$ $(-1)$ the polarization $a$, $(a\perp)$, and by $Y = +1$ $(-1)$ the polarization $b$, $(b\perp)$. The following probabilities will be useful in the analysis below

(4) $P[X = +1] = P[X = +1] = \frac{1}{2}$,

as results from (3). Using the equality $P[Y|X] = P[XY]/P[X]$ and (4), we have from (3)

(5) $\begin{cases} P[Y = +1|_{\theta,\phi,X=+1}] = \cos^2(\phi - \theta) & P[Y = -1|_{\theta,\phi,X=+1}] = \sin^2(\phi - \theta) \quad\text{(a)}, \\ P[Y = +1|_{\theta,\phi,X=-1}] = \sin^2(\phi - \theta) & P[Y = -1|_{\theta,\phi,X=-1}] = \cos^2(\phi - \theta) \quad\text{(b)}. \end{cases}$

Colbeck and Renner [1] describe the results $X$ and $Y$ as follows:

$X = f(A, B, U, V, W), \quad Y = g(A, B, U, V, W)$,

where $A$, $B$, are the settings at the station of Alice, respectively Bob, in our case $A = \theta$, $B = \phi$, $U$ and $V$ are the respective local hidden variables (HVs), and $W$ is the non-local HV. $U$, $V$, and $W$, are assumed independent of one another. According to the definition of local HV, $U$ should not influence the result $Y$ and $V$ should not influence the result $X$, s.t. the functions $g$ and $f$ become

(6) $X = f(\theta, \phi, U, W), \quad Y = g(\theta, \phi, V, W)$.

A special assumption is made in [1] on the conditional probabilities $P[X|_{\theta,\phi,U,W}]$ and $P[Y|_{\theta,\phi,V,W}]$, namely that averaging over $W$ eliminates the dependence of $P[X|_{\theta,\phi,U,W}]$ on $\theta$, and of $P[Y|_{\theta,\phi,V,W}]$ on $\phi$:

(7) $\sum_w P_W(w)\, P[X|_{\theta,\phi,U=u,W=w}] = P[X|_{\theta,U}]$ $\quad$ (a),

$\sum_w P_W(w)\, P[Y|_{\theta,\phi,V=v,W=w}] = P[Y|_{\phi,V}]$ $\quad$ (b).

To see what that means let's introduce in (7) the equalities $P[X|U] = P[UX]/P[U]$, and $P[Y|V] = P[VY]/P[V]$,

(8) $\sum_w P_W(w)\, P[(U=u)\&(X=x|_{\theta,\phi,W=w})] = P_U(u)\, P[X|_{\theta,U}]$ $\quad$ (a),

$\sum_w P_W(w)\, P[(V=v)\&(Y=y|_{\theta,\phi,W=w})] = P_V(v)\, P[Y|_{\phi,V}]$ $\quad$ (b).

(8a) states that the joint probability of $U$ and $X$ doesn't depend on $\phi$, and (8b) states that the joint probability of $V$ and $Y$ doesn't depend on $\theta$. That resembles the non-signaling property, which says that the probabilities of the values of an *observable* of a particle do not depend on the settings at the station where the other particle is tested. But in the way $U$ and $V$ are used in [1], see [3], they are not observables according to the lemma in section 1, so neither are the pairs $U\&X$ and $V\&Y$ in (8). Thus, the assumptions (7) are not justified.

## 4. A counter-example

The fact that the Colbeck and Renner class of models does not permit local HVs was proved in [2] without applying to the assumptions (7). But this is not the last word about local HVs in combined models. Other types of combined models than in [1] may be built, as exemplified below.

Assume that $X$ depends only on the local data $\theta$ and $U$, and that $Y$ depends on both the local data $\phi$ and $V$ and the non-local data $\theta$ and $X$

(9) $X = f(\theta, U), \quad Y = g(\theta, \phi, V, X)$.

This model *does not* belong to the class studied in [1]. Comparing with (6), there are no non-local HVs here. In fact, it is not known what is a non-local HV and this concept is not at all clear. Instead of a non-local HV, the function $g$ comprises the stochastic, non-local parameter $X$, which is an observable, but Bob is unaware of its value.
In [1] $U$, $V$, and $W$ are mutually independent, while in (9) the non-local parameter $X$ of $g$ depends on $U$ through the function $f$.
But the main difference is that (9) offers an example against the assumption (7b), as shown below. Introducing the probabilities (4), (7b) takes the form

(10) $½\{P[Y|_{\theta, \phi, V=v, X=+1}] + P[Y|_{\theta, \phi, V=v, X=-1}]\} = P[Y|_{\phi, V=v}]$.

For shortening the discourse, the expression on the L.H.S. of (10) will be denoted by $L$.
Let $\{v_1\}$ be the set of values of $V$ for which $g(\theta, \phi, v_1, X=+1) = +1$, and $\{v_2\}$ the set of values of $V$ for which $g(\theta, \phi, v_2, X=-1) = +1$, fig. 1. Thus, $\{v_1\} \cup \{v_2\}$ produces all the cases of $Y=+1$.
Fig. 1a shows a distribution with $\{v_1\}$ and $\{v_2\}$ disjoint for any $\phi$ and $\theta$. Since according to (5) the total probability of the set $\{v_1\}$ is $\cos^2(\phi - \theta)$, and of $\{v_2\}$ is $\sin^2(\phi - \theta)$, any value $v$ of $V$ belongs necessarily to one of these two sets, and only to one. Then, for $Y=+1$ and whatever $v$, one gets $L = ½$. By an analogous rationale, for $Y=-1$ and whatever $v$, one gets also $L = ½$. Thus, $L$ does not depend on $\theta$ and (10) is correct.
Not so is the situation if $\{v_1\}$ and $\{v_2\}$ are not disjoint.
Let's denote $\{v_\cap\} = \{v_1\} \cap \{v_2\}$. Fig. 1b shows the case of maximal overlapping of $\{v_1\}$ and $\{v_2\}$. Three situations may be distinguished for an arbitrary value $v$ of $V$: a) $v \in \{v_\cap\}$, which implies $L = 1$; b) $v \in \{v_1\}-\{v_\cap\}$, or, $v \in \{v_2\}-\{v_\cap\}$, which implies $L = ½$; c) $v$ doesn't belonging to either $\{v_1\}$ or $\{v_2\}$, which implies $L = 0$.
The total probability of $\{v_\cap\}$ is $\sin^2(\phi - \theta)$, which means that the number of values in $\{v_\cap\}$ varies with $\phi$ and $\theta$. Let's remind that $\{v_\cap\}$ comprises for given $\phi$ and $\theta$ all the values of $V$ that produce $Y=+1$ for both $X=\pm1$. The probability of occurrence of $v$ does not depend on $\theta$; however whether $v$ produces $Y=+1$ or $Y=-1$, for $X=+1$ and/or for $X=-1$, that does depend on $\theta$, see (9). Thus, there are angles $\theta$ for which $v$ falls in the set $\{v_\cap\}$, in which case $L = 1$, and there are angles $\theta$ for which $v$ falls outside $\{v_\cap\}$, e.g. only in $\{v_1\}$, in which case $L = ½$. Therefore for this $v$, $L$ depends on $\theta$.
In conclusion, if $\{v_1\}$ and $\{v_2\}$ are not disjoint (10) is not correct, and that invalidates (7b).

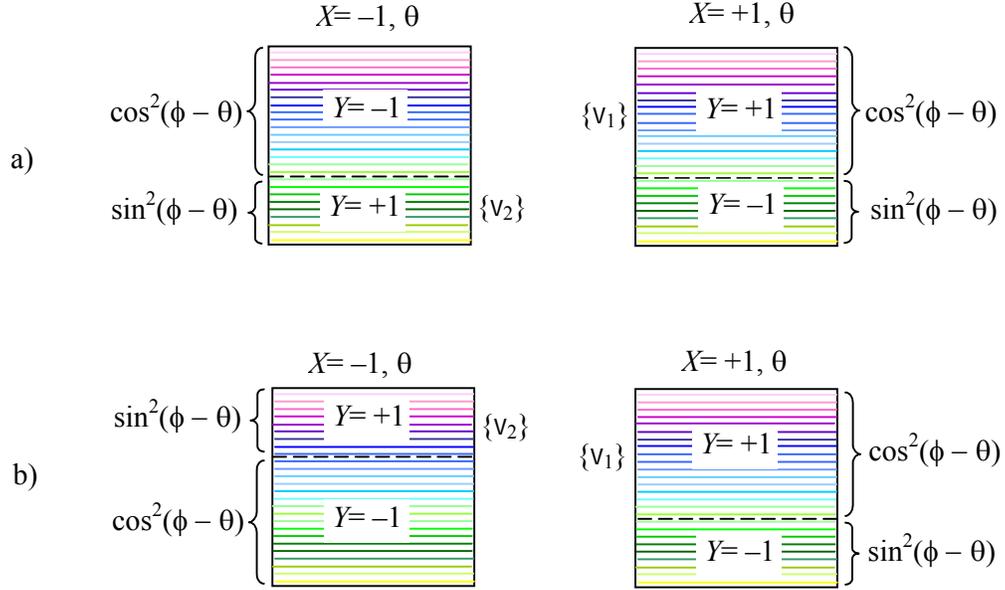

**Figure 1.** Two possible distributions of the local hidden variable $V$.
Each square contains all the values of $V$, each value being represented by a line of some color. A square is partitioned into two sets according to the result $Y$ produced, and on the side of each set appears the total probability of the values in it. Above each square are listed the non-local data.
a) the sets that produce the same value of $Y$ are complementary; b) the sets that produce the same value of $Y$ are not disjoint.

## 5. Discussion

The example in the previous section places the assumptions made in [1] under question mark, and so their conclusion.

Actually, the conclusion in [1] is that $P_{XU|\theta} = \mathcal{U}_X \times P_U$, and $P_{YV|\phi} = \mathcal{U}_Y \times P_V$, where $\mathcal{U}_X, \mathcal{U}_Y$, are the uniform distributions of the values of $X$, respectively $Y$. This is exactly the situation in fig. 1a, where each value of $V$ generates with equal probability $Y = +1$ and $Y = -1$. That doesn't mean that local HVs do nothing.

The question whether HVs exist or not, i.e. whether the microscopic world is deterministic or really non-deterministic – "God plays dice" – remains open, all the more that the HVs are not observables.